\documentclass[letterpaper]{article}

\usepackage[letterpaper]{geometry}
\usepackage{amsmath}
\usepackage{amsfonts}
\usepackage{relsize}
\usepackage{braket} %braket, bra, ket
\usepackage{hyperref}
\usepackage{xcolor}

\usepackage{siunitx}
\definecolor{LightGray}{gray}{0.9} %background color for python syntax highlight
\usepackage[frozencache,cachedir=.]{minted}%for python syntax highlighting
\setminted{linenos, framesep=2mm, mathescape, autogobble, baselinestretch=1.2,
           bgcolor=LightGray, fontsize=\small, frame=lines, framerule=0.5 pt}
\newcommand{\setsd}{\ensuremath S_{\mathbf{m}}}

\usepackage[backend=biber,style=chem-acs]{biblatex}
\defbibheading{bibliography}[\bibname]{\section{References}}
\title{\texttt{Fanpy}: A Python Library for Prototyping
  Multideterminant Methods in \textit{Ab Initio} Quantum Chemistry}

\author{Taewon D. Kim$^1$
  \and Michael Richer$^1$
  \and Gabriela S\'{a}nchez-D\'{i}az$^1$
  \and Farnaz Heidar-Zadeh$^2$
  \and Toon Verstraelen$^3$
  \and Ram\'{o}n Alain Miranda-Quintana$^4$
  \and Paul W. Ayers$^1$}
\date{\today}

\begin{document}
\maketitle
\vspace*{-1.5em}
\noindent
$^1$Department of Chemistry and Chemical Biology, McMaster University, Hamilton,
Ontario, L8S-4L8, Canada\\
$^2$Department of Chemistry, Queen's University, Kingston, Ontario, K7L-3N6,
Canada\\
$^3$Center for Molecular Modeling (CMM), Ghent University, Technologiepark-Zwijnaarde
46, B-9052, Zwijnaarde, Belgium\\
$^4$Department of Chemistry, University of Florida, Gainesville, FL 32603, USA\\

\begin{abstract}
  \texttt{Fanpy} is a free and open-source Python library for developing and
  testing multideterminant wavefunctions and related \textit{ab initio} methods
  in electronic structure theory.
  The main use of \texttt{Fanpy} is to quickly prototype new methods by
 making it easier to transfer the mathematical conception of a new wavefunction ans\"{a}tze to a working implementation. \texttt{Fanpy} uses the framework of our recently introduced Flexible Ansatz for N-electron Configuration Interaction (FANCI), where multideterminant
  wavefunctions are represented by their overlaps with Slater determinants of orthonormal spin-orbitals. In the simplest case, a new wavefunction ansatz can be implemented by simply writing a function for evaluating its overlap with an arbitrary Slater determinant. \texttt{Fanpy} is modular in both implementation and theory: the wavefunction model, the system's Hamiltonian, and the choice of objective function are all independent
  modules. This modular structure makes it easy for users to mix and match different methods and for developers to quickly try new ideas. \texttt{Fanpy} is written purely in Python with standard dependencies, making
  it accessible for most operating systems; it adheres to principles of modern software development, including comprehensive documentation, extensive testing, and continuous integration and delivery protocols.
  This article is considered to be the official release notes for the \texttt{Fanpy} library.

% Please include a maximum of seven keywords
% \keywords{quantum chemistry, electronic structure theory, computational
%   chemistry, ab initio methods, theoretical chemistry Python library}
\end{abstract}

\section{What is \texttt{Fanpy}?}
\texttt{Fanpy} is a free and open-source Python 3 library for \textit{ab initio}
electronic structure calculations. The key innovation is the adoption of the Flexible Ansatz for N-electron Configuration Interaction (FANCI) mathematical framework\cite{fanci}. By adopting this framework, \textit{ab initio} electronic structure methods are represented as a collection of four parts, each of which is represented by an independent module of \texttt{Fanpy}:
(a) the (multideterminant) wavefunction model
(b) the system Hamiltonian, as represented by its one- and two-electron integrals
(c) an equation (or system of equations) to solve that is equivalent to the Schr\"{o}dinger equation,
(d) an algorithm for optimizing the objective function(s).
Section \ref{sec:fanpy_features} details the features of each module, though the main
advantage of \texttt{Fanpy} is that new methods can be implemented easily.

Although the FANCI framework allows overlaps with any convenient set of reference states (e.g., Richardson eigenfunctions\cite{apr2g}) to be used, in \texttt{Fanpy} the aforementioned components are expressed explicitly in terms
of Slater determinants and the wavefunction ans\"{a}tze of interest are multideterminant wavefunctions with parameterized coefficients,
\begin{equation}
  \label{eq:fanpy_FANCI_fanpy}
  \ket{\Psi_{\mathrm{FANCI}}} = \sum_{\mathbf{m} \in \setsd} f(\mathbf{m}, \vec{P}) \ket{\mathbf{m}}
\end{equation}
where
\begin{equation}
  \begin{split}
    \ket{\mathbf{m}} = \ket{m_1 m_2 \dots m_{N-1} m_N} = a^\dagger_{m_1} a^\dagger_{m_2} \dots a^\dagger_{m_{N-1}} a^\dagger_{m_N} \ket{}\\
  \end{split}
\end{equation}
denotes a Slater determinant, $\setsd$ is the set of Slater determinants included in the wavefunction, and $f$ is a function that determines the coefficient of each Slater determinant, $\mathbf{m}$, using the parameters, $\vec{P}$. Note that $f$ is simply the overlap of the parameterized wavefunction with the Slater determinant,
\begin{equation}
  \label{eq:fanpy_overlap_fanpy}
  f(\mathbf{m}, \vec{P})= \braket{\mathbf{m} | \Psi_{\mathrm{FANCI}}}
\end{equation}
Similarly, the Hamiltonian is expressed in terms of its CI matrix elements,
\begin{equation}
  \label{eq:fanpy_Hamiltonianmatrix_fanpy}
  \braket{\mathbf{m} | \hat{H} | \mathbf{n}} = \braket{\mathbf{m} |  \sum_{ij} h_{ij} a^\dagger_i a_j
    + \frac{1}{2} \sum_{ijkl} g_{ijkl} a^\dagger_i a^\dagger_j a_l a_k | \mathbf{n}}
\end{equation}
The objective functions supported by \texttt{Fanpy} combine the overlaps and the CI matrix elements to approximate the Schr\"{o}dinger equation.

\section{About \texttt{Fanpy}}
The source code of \texttt{Fanpy} is maintained on GitHub; see
\url{https://github.com/quantumelephant/fanpy}, and its documentation is hosted
on Read the Docs; see \url{https://fanpy.readthedocs.io/en/latest/index.html}.
We strive to ensure that the \texttt{Fanpy} source code and its associated website are
comprehensively documented, including useful tests, scripts, and examples.
As that documentation is maintained with the software, providing detailed (and
eventually outdated) release notes here seems unwise.
Instead, we will briefly list the distinguishing features and key capabilities
of \texttt{Fanpy} in section \ref{sec:fanpy_features} to establish its philosophy and
framework, and exemplify them in section \ref{sec:fanpy_examples}.

\section{Why \texttt{Fanpy}?}
% existing packages
% Many proprietary quantum chemistry software provides source codes,
% such as Guassian, Molpro, Orca, and Gamess
% Though open-source, lacking documentation and support for developers
% Developer focused, psi4numpy and pyquante,
Many quantum chemistry packages enable computations using multideterminant methods. Most of these packages (e.g.
Gaussian\cite{gaussian} and MolPro\cite{molpro}) are closed-source, making it nearly impossible to
develop new methods without special permission. Even for packages whose source code is available, the code is often monolithic, making it difficult to implement new fundamental methods without thoroughly understanding nearly the entire code base. Moreover, the low-level code is often highly optimized, abstracts away critical components of \textit{ab initio} methods, and not intended to be subsequently read or modified. Such code can, and does, remain unchanged for decades, has little documentation, and rarely follows modern software development principles. Though some packages try to address this issue, the development of post-HF methods
remains difficult. For example, Psi4Numpy\cite{psi4numpy} is a collection of Python scripts and
Jupyter notebooks that implement several post-HF methods using Psi4 to generate
necessary inputs, such as CI matrices and one- and two-electron integrals.
Though these scripts are excellent tools for learning about standard quantum chemistry methods or implementing minor embellishments of standard methods, developing
a novel method (e.g. a new wavefunction ansatz) requires all related processes to be
implemented in Psi4, which requires a thorough understanding of the Psi4 package as a whole\cite{psi4}. In particular, implementing a new objective function and its associated optimization algorithm, or implementing the action of the Hamiltonian on a new wavefunction ansatz, is difficult and time-consuming.

Due to the difficulty of developing new methods in these legacy codes, we developed our own Helpful Open-Source Research TOol for N-electron systems (HORTON)\cite{horton}. The first two versions of HORTON were monolithic, and wavefunction models that were simple on paper were difficult to implement. HORTON 3 is strictly modular, with separate modules for input/output, numerical integration, Gaussian-basis-set evaluation and integrals, geometry optimization, and self-consistent field calculations. \texttt{Fanpy} is the correlated wavefunction module of HORTON 3.
%many
%modern quantum chemistry packages, such as HORTON\cite{horton},
%PySCF\cite{pyscf}, and Psi4\cite{psi4}, were developed with consideration
%towards ease of development and maintainability.
% often starting off as an software within a research group that could then be
% inherited and extend upon by the next generation of graduate students.

\texttt{Fanpy} was envisioned as a development tool for new correlated-wavefunction methods; the goal is to help researchers quickly implement and test their ideas. Towards this goal, \texttt{Fanpy} is designed to be modular and general. Its modularity helps to isolate and minimize the amount of code that needs to be understood, and perhaps modified, to implement a new method. For example, implementing a new wavefunction ansatz requires modifying only the wavefunction module, and does not require explicit consideration of how the Hamiltonian will act upon that wavefunction nor of how the orbitals and parameters within the wavefunction will be optimized. The modules of \texttt{Fanpy} are designed to be as general as possible, so that features from one module are compatible with features from the other modules. The compatibility between the modules ensures that any developed method (e.g. a wavefunction ansatz) can be used in conjunction with the other methods (e.g. orbital optimization, model Hamiltonians, the projected Schr\"{o}dinger equation, etc).
% Each wavefunction can be combined with each Hamiltonian to form different
% objectives corresponding to the Schr\"{o}dinger equation, which can then be
% solved by a compatible optimization algorithm.
We provide comprehensive documentation and examples to further aid the
development of new methods in \texttt{Fanpy}.
% Several structures have been put in place to ensure that the developer produces
% mingles well with the other modules.

\section{Features of \texttt{Fanpy}}
\label{sec:fanpy_features}
We display various features of \texttt{Fanpy} by discussing each module and
their intended purposes:

% \begin{itemize}[topsep=4pt, itemsep=4pt, partopsep=4pt, parsep=4pt]
\begin{itemize}
\item The wavefunction module is developed in accordance with the FANCI framework\cite{fanci}.
  In the FANCI framework, the wavefunction is entirely represented by its
  overlaps with Slater determinants built from orthonormal orbitals.
  Similarly, each wavefunction in \texttt{Fanpy} is defined by its parameters
  and a function that returns an overlap for the given Slater determinant.
  The overlap can be provided as a standalone function or defined within a class structure, templated from an abstract base class. The following wavefunctions have already been implemented: configuration interaction (CI) with single and double excitations (CISD)\cite{cisd}; doubly-occupied configuration interaction (DOCI)\cite{7, 10, 15, doci}; full CI\cite{fci}; selected CI wavefunctions with a user-specified set of Slater determinants; antisymmetrized products of geminals (APG)\cite{39, 40, 41, 42, 43, 44, 45, 46, 47, geminal_first, apg}; antisymmetrized products of geminals with disjoint orbital sets (APsetG)\cite{73}; antisymmetrized product of interacting geminals (APIG)\cite{48, 49, 50, 51, 52, 53, 54, 55, 56, 57, 58, 59, 60, 61,
    62, 63, 64, agp, 66, 67, 68, 69, 70, 71, 72, 73, 74, 75, apig_first}; antisymmetric product of 1-reference-orbital interacting geminals (AP1roG; equivalent to pair-coupled-cluster doubles)\cite{ap1rog}; antisymmetric product of rank-two interacting geminals (APr2G)\cite{apr2g}; determinant ratio wavefunctions\cite{fanci}; antisymmetrized products of tetrets (4-electron wavefunctions)\cite{fanci}; matrix product states (MPS)\cite{mps_dmrg}; neural network wavefunctions; coupled-cluster (CC) with arbitrary excitations (including, but not limited to, CCSD, CCSDT, and CC with seniority-specific excitations)\cite{22, 23, 24, cc_review, ramonemail_cc1, ramonemail_cc2,fanci}, geminal coupled-cluster wavefunctions\cite{55, 56, 57, 59, ap1rog}, generalized CC, and seniority-increasing CC. We also support these wavefunctions with nonorthogonal orbitals, and linear combinations of any of the aforementioned wavefunctions.

\item The Hamiltonian module contains Hamiltonians commonly used in electronic
  structure theory. Similar to the wavefunctions, each Hamiltonian in \texttt{Fanpy} is defined by
  its representation in orbital basis set (i.e. one- and two-electron integrals)
  and a function that returns the integral of the Hamiltonian with respect to
  the given Slater determinants. The following Hamiltonians have already been implemented: the electronic Hamiltonian in the restricted, unrestricted, and generalized basis; the seniority-zero electronic Hamiltonian\cite{sen0_ham}; and the Fock operator in the restricted basis. In addition, the Pariser-Parr-Pople\cite{ppp1,ppp2,ppp3,huckel}, Hubbard\cite{hubbard_ham,huckel}, H\"{u}ckel\cite{huckel,huckel2}, Ising\cite{ising,models,huckel},
  Heisenberg\cite{heisenberg,models,huckel}, and Richardson\cite{richardson1,richardson2} model Hamiltonians are available as restricted electronic
  Hamiltonians through the ModelHamiltonian GitHub repository\cite{modelhamiltonian}.
  Orbital optimization is available if a function returning the derivative with respect to orbital
  rotation parameters is provided. At the moment, only restricted electronic Hamiltonians 
  support orbital optimization.

\item The objective module is responsible for combining the wavefunction and the
  Hamiltonian to form an equation or a system of equations that represents the
  Schr\"{o}dinger equation. In \texttt{Fanpy}, the objective function can be the variational optimization of the expectation value of the energy\cite{piela,jensen,cramer,1,szabo}, the projected Schr\"{o}dinger equation\cite{59,72,73}, or a local energy expression to be sampled (as in variational quantum Monte Carlo)\cite{qmc2,qmc,vqmc,vqmc4,vqmc7,vqmc8}.

  %Schr\"{o}dinger equation can be represented as energy
  %(Equation~\ref{eq:fanpy_energy}) used in variational methods,
  % local energy
  % (Equation~\ref{eq:fanpy_local_energy}) used in orbital space variational Quantum
  % Monte Carlo,
  %\textcolor{blue}{\textbf{[NOTE: include local energy? (in example later)]}}
  %and Projected Schr\"{o}dinger equation (Equation~\ref{eq:fanpy_projected}) often
  %used to solve geminal wavefunctions.
  %\begin{equation}
  %  \label{eq:fanpy_energy}
  %  E = \frac{\braket{\Psi | hat{H} | \Psi}}{\braket{\Psi | \Psi}}
  %\end{equation}
  % \begin{equation}
  %   \label{eq:fanpy_local_energy}
  %   E = \sum_i \frac{\braket{\Phi_i | hat{H} | \Psi}}{{\braket{\Phi_i | \Psi}}}
  % \end{equation}
  % \begin{equation}
  %   \label{eq:fanpy_projected}
  %   \begin{split}
  %     \braket{\Phi_1 | hat{H} | \Psi} - E \braket{\Phi_1 | \Psi} &= 0\\
  %     &\vots\\
  %     \braket{\Phi_M | hat{H} | \Psi} - E \braket{\Phi_M | \Psi} &= 0\\
  %   \end{split}
  % \end{equation}
  %where $\Phi_i$ is a Slater determinant.

\item The solver module contains algorithms that optimize/solve the equations from the
  objective module. It supports optimizers from \texttt{SciPy}\cite{scipy}, which includes
  constrained/unconstrained local/global optimizers for multivariate scalar
  functions (i.e. energy) and algorithms for solving nonlinear least-squares
  problems and for finding roots of a system of nonlinear equations (i.e.,
  projected Schr\"{o}dinger equation). For CI wavefunctions, we also support brute-force eigenvalue
  decomposition. In addition, \texttt{Fanpy} interfaces to several algorithms for  derivative-free global optimization problems including the Covariance Matrix Adaptation Evolution Strategy (CMA-ES) algorithm\cite{cma} from \texttt{pycma}\cite{pycma} and algorithms using decision trees and Bayesian optimization from \texttt{scikit-optimize}\cite{scikit}. At the moment, no in-house optimization algorithms specialized for
  electronic structure theory problems are included. However, \texttt{Fanpy}'s modular design makes it easier to develop sophisticated domain-specific optimization algorithms. The objective module provides high-level control over the parameters involved in the optimization (e.g., active and frozen parameters) and can be changed dynamically
  throughout the optimization process.
  These parameters can be saved as a checkpoint throughout the
  optimization. (The default is to checkpoint at each function evaluation.)
  Furthermore, the objective module provides flexibility to add additional
  parameters (e.g., model hyperparameters) and to add nonlinear constraints
  to the Projected Schr\"{o}dinger equation.

\item The \texttt{tool} module provides various utility functions used throughout the
  \texttt{Fanpy} package.
  Though some tools have specialized uses, the tools for manipulating and
  generating Slater determinants are used frequently throughout \texttt{Fanpy}.
  These tools are essential when developing methods in \texttt{Fanpy} because
  Slater determinants are the common language of the independent modules.
  The \texttt{slater} module provides functions for manipulating Slater
  determinants and converting them from one form to another.
  Within \texttt{Fanpy}, Slater determinants are represented as a binary number,
  where the positions of 1's are the indices of the occupied spin-orbitals.
  The \texttt{slater} module can, for example, provide the occupied spin orbital
  indices from the given Slater determinant.
  The \texttt{sdlist} module provides easy ways to generate Slater determinants
  of the desired characteristics (e.g. order of excitation from ground state,
  spin, seniority).
  This module is frequently used to construct the projection space by which the objective function is evaluated.
  In addition, the \texttt{tool} module provides wrappers to other modules of \texttt{HORTON} and other quantum
  chemistry software, including \texttt{Gaussian}\cite{gaussian},
 \texttt{PySCF}\cite{pyscf}, and
  \texttt{Psi4}\cite{psi4}. These programs can then be used to generate one- and
  two-electron integrals for \texttt{Fanpy}.

\end{itemize}

\section{Examples} \label{sec:fanpy_examples}

For the most updated documentation and examples on how to use \texttt{Fanpy},
please refer to the \texttt{Fanpy} website.
Here, we showcase several ways \texttt{Fanpy} can be used and incorporated into
various workflows.
Please note that these examples are based on version 1.0 of \texttt{Fanpy}, and
the user might need to modify them if using a future major release of the
\texttt{Fanpy} library.
Within minor and bug-fix releases, backward compatibility is guaranteed.

\subparagraph{Running a calculation:}
A calculation in \texttt{Fanpy} can be run by creating and executing a Python
script and by running its command-line tool, \texttt{fanpy\char`_run\char`_calc}.
For ease of use, \texttt{fanpy\char`_run\char`_calc} provides limited access to \texttt{Fanpy}'s features using sensible default settings.
However, it is recommended to create and execute a Python script because it provides a
transparent record of the calculation (and its settings) and the full range of \texttt{Fanpy}'s
features.
For assistance in creating a Python script, \texttt{Fanpy} provides a
command-line tool, \texttt{fanpy\char`_make\char`_script}.
This tool creates a script from the given specifications, which can then be
modified if a desired feature is not available in the tool.

The following example of a Python script runs an AP1roG calculation for
oxygen molecule in a double zeta basis set:
% check if we can actually run benzene
% check number of spin orbitals in double zeta benzene
\begin{minted}{python}
import numpy as np
from fanpy.wfn.geminal.ap1rog import AP1roG
from fanpy.ham.restricted_chemical import RestrictedMolecularHamiltonian
from fanpy.eqn.projected import ProjectedSchrodinger
from fanpy.solver.system import least_squares
from fanpy.tools.sd_list import sd_list

nelec = 16

# Hamiltonian
oneint = np.load('one_oxygen.npy')
twoint = np.load('two_oxygen.npy')
ham = RestrictedMolecularHamiltonian(oneint, twoint)

# Wavefunction
wfn = AP1roG(nelec, ham.nspin)

# Projection space of first and second order excitation
pspace = sd_list(nelec, ham.nspin, exc_orders=[2], seniority=0)

# Projected Schrodinger Equation
eqns = ProjectedSchrodinger(wfn, ham, pspace=pspace)

# Solve
results = least_squares(eqns)
print('AP1roG electronic energy (Hartree):', results['energy'])
\end{minted}

Since \texttt{Fanpy} targets post-HF methods, the orbitals (and the
corresponding system specific information) must be provided in the form of one-
and two-electron integrals.
The one- and two-electron integrals must be provided as two- and
four-dimensional numpy arrays, respectively, whose indices are in the same order
as the integrals in the physicists' notation.
To generate the integrals from a single-determinant calculation, \texttt{Fanpy}
provides wrappers for \texttt{HORTON}, \texttt{PySCF}, and \texttt{Psi4} via the
\texttt{fanpy.tools.wrapper} module.
The Gaussian \texttt{.fchk} file can be converted into \texttt{.npy} file using
the HORTON wrapper, which will also compute the required one- and two-electron integrals.

\subparagraph{Implementing a wavefunction:}
New wavefunctions can be implemented in \texttt{Fanpy} by making a subclass of
the wavefunction base class or by providing the overlap function to a utility
function. The subclass requires the method \texttt{get\char`_overlap} to be defined.

As a simple example, recall that expansion of a Slater determinant of nonorthogonal orbitals in orthogonal Slater determinants is given by:
\begin{equation}
  \label{eq:fanpy_nonorth}
  \begin{split}
    \ket{\Psi} &= \prod_{i=1}^N \sum_{j}^{2K} C_{ij} a^{\dagger}_j \ket{\theta}\\
    &= \sum_{\mathbf{m}} |C(\mathbf{m})|^- \ket{\mathbf{m}}\\
  \end{split}
\end{equation}
In \texttt{Fanpy}, this corresponds to:
\begin{minted}{python}
import numpy as np
from fanpy.wfn.base import BaseWavefunction
from fanpy.tools.slater import occ_indices

class NonorthogonalSlaterDeterminant(BaseWavefunction):
    def get_overlap(self, sd, deriv=None):
        # get indices of the occupied spin orbitals
        occs = occ_indices(sd)
        # reshape the parameters
        # NOTE: parameters are stored as a one-dimensional array by default
        params = self.params.reshape(self.nelec, self.nspin)
        # compute the overlap
        if deriv is None:
            return np.linalg.det(params[:, occs])

        # compute the derivative of the overlap
        output = np.zeros(params.shape)
        for deriv_row in range(self.nelec):
            for j, deriv_col in enumerate(occs):
                # compute the sign associated with Laplace formula
                sign = (-1)**(deriv_row + j)
                # get rows and columns with the appropriate row/column removed
                row_inds = np.arange(self.nelec)
                row_inds = row_inds[row_inds != deriv_row]
                col_inds = occs[occs != deriv_col]
                # compute minors (determinant after removing row and column)
                minor = np.linalg.det(params[row_inds[:, None], col_inds[None, :]])
                output[deriv_row, deriv_col] = sign * minor
        # derivative is returned as a flattened array
        # deriv contains the indices of the parameters with respect to which
        # the overlap is derivatized
        return output.ravel()[deriv]
\end{minted}
The method \texttt{get\char`_overlap} returns the overlap of the given Slater
determinant when \texttt{deriv=None} and returns its gradient with respect to
the parameters specified by \texttt{deriv} otherwise (\texttt{deriv} is a
one-dimensional numpy array of parameter indices).
Further details on the API of \texttt{get\char`_overlap} are provided in the online documentation.
Unlike the wavefunctions already implemented in \texttt{Fanpy}, this
wavefunction does not have default initial parameters, which means that they
must be supplied when instantiating the wavefunction. For example, the following code block shows how to initialize to the ground-state (orthogonal) Slater determinant.
\begin{minted}{python}
from fanpy.tools.slater import ground, occ_indices

# get indices of the HF ground state
ground_indices = occ_indices(ground(16, 36))
# initial parameters (only contain the occupied orbitals in HF ground state)
hf_params = np.zeros((16, 36))
hf_params[np.arange(16), ground_indices] = 1

wfn = NonorthogonalSlaterDeterminant(16, 36)
# assign parameters
wfn.assign_params(hf_params)
\end{minted}

Alternatively, the wavefunction can be constructed using the utility function:
\texttt{wfn\char`_factory}.
\begin{minted}{python}
import numpy as np
from fanpy.wfn.utils import wfn_factory

def olp(sd, params):
    occs = occ_indices(sd)
    # NOTE: Since the only information available come from the arguments sd and
    # params, additional information that would otherwise be stored as
    # instance's attributes and properties must be explicitly defined
    nelc = 16
    nspin = 36
    # reshape the parameters
    params = params.reshape(nelec, nspin)
    return np.linalg.det(params[:, occs])

def olp_deriv(sd, params):
    occs = occ_indices(sd)
    # hardcode essential information
    nelc = 16
    nspin = 36
    # reshape the parameters
    params = params.reshape(nelec, nspin)
    # same as above except replace self.nelec with nelec
    # ...
    # NOTE: the overlap is derivatized with respect to all wavefunction
    # parameters unlike above
    return output.ravel()

wfn = wfn_factory(olp, olp_deriv, 16, 36, hf_params)
# third argument is the number of electrons
# fourth argument is the number of spin orbitals
# fifth argument is the initial parameters
\end{minted}

It is recommended to implement wavefunctions using the class structure because
it helps make the code cleaner by limiting repetitions and makes the code easier to unit test
by breaking it into smaller pieces.
For a quick and dirty implementation, however, the utility function may be
easier.

\subparagraph{Implementing a Hamiltonian:}
Similar to the wavefunction, new Hamiltonians can be implemented in
\texttt{Fanpy} by making a subclass of the Hamiltonian base class or by
passing a function that evaluates the integrals to a utility function.
In addition to the general Hamiltonian base class, \texttt{Fanpy} provides base
classes according to the type of orbitals used in the Hamiltonian: restricted,
unrestricted, and generalized.
The subclass to the orbital specific Hamiltonian base class requires the
method \texttt{integrate\char`_sd\char`_sd} to be defined.

For example, to implement the H\"{u}ckel Hamiltonian:\cite{huckel}
\begin{equation}
  \label{eq:fanpy_huckel}
  \begin{split}
    \hat{H} &= \sum_{ij} \sum_{\sigma} h_{ij} a^\dagger_{i\sigma} a_{j\sigma}\\
    h_{ij} &=
    \begin{cases}
      \alpha_i &\mbox{if $i = j$}\\
      \beta_{ij} &\mbox{if spatial orbitals $i$ and $j$ belong to atoms that participate in a bond}\\
      0 &\mbox{else}
    \end{cases}
  \end{split}
\end{equation}
\begin{minted}{python}
from fanpy.ham.base import BaseHamiltonian
from fanpy.tools import slater

class HuckelHamiltonian(BaseHamiltonian):
    def __init__(self, one_int):
        # NOTE: provided integrals correspond to spatial orbitals
        self.one_int = one_int

    @property
    def nspin(self):
        return self.one_int.shape[0] * 2

    def integrate_sd_wfn(self, wfn, sd, wfn_deriv=None, ham_deriv=None):
        # use the default method except only the first order excitations are used
        return super().integrate_sd_wfn(
            wfn, sd, wfn_deriv=wfn_deriv, ham_deriv=ham_deriv, orders=(1,)
        )

    def integrate_sd_sd(self, sd1, sd2, deriv=None, components=False):
        # get the difference of the Slater determinants (i.e. which orbitals are
        # occupied in one determinant but not in the other)
        diff_sd1, diff_sd2 = slater.diff_orbs(sd1, sd2)
        # derivative not supported here
        if deriv:
            raise NotImplementedError
        # if order of excitation between the two Slater determinants is two or greater
        if len(diff_sd1) >= 2 or len(diff_sd2) >= 2:
            return 0.0
        # if two Slater determinants do not have the same number of electrons
        if len(diff_sd1) != len(diff_sd2):
            return 0.0
        # if two Slater determinants are the same
        if len(diff_sd1) == 0:
            # get the indices of the spatial orbitals that correspond to the
            # occupied spin orbitals
            shared_alpha_sd, shared_beta_sd = slater.split_spin(
                slater.shared_sd(sd1, sd2), self.nspatial
            )
            shared_alpha = slater.occ_indices(shared_alpha_sd)
            shared_beta = slater.occ_indices(shared_beta_sd)
            # sum over the occupied orbitals
            output = np.sum(self.one_int[shared_alpha, shared_alpha])
            output += np.sum(self.one_int[shared_beta, shared_beta])
            return output
        # if two Slater determinants are different by one-electron excitation
        # get indices of the spatial orbitals
        spatial_ind1 = slater.spatial_index(diff_sd1[0], self.nspatial)
        spatial_ind2 = slater.spatial_index(diff_sd2[0], self.nspatial)
        return self.one_int[spatial_ind1, spatial_ind2]
\end{minted}
Though it is not necessary, the subclass defines \texttt{integrate\char`_sd\char`_wfn} to
specify that the Hamiltonian only contains one-body operators.
By default, \texttt{integrate\char`_sd\char`_wfn} assumes that the Hamiltonian contains one-
and two-body operators.

Alternatively, the hamiltonian can be constructed using the utility function
\begin{minted}{python}
from fanpy.ham.utils.factory import ham_factory
from fanpy.tools import slater

def integrate_sd_sd(sd1, sd2, one_int):
    diff_sd1, diff_sd2 = slater.diff_orbs(sd1, sd2)
    nspatial = one_int.shape[0]
    # same as above except replace self.one_int with one_int
    # and self.nspatial with nspatial
    # ...

ham = ham_factory(integrate_sd_sd, oneint, 36, orders=(1,))
# third argument is the number of electrons
\end{minted}
Again, using the class structure is encouraged because its structure can be
cleaner and more transparent and because it provides finer control over the
Hamiltonian.
For example, if \texttt{integrate\char`_sd\char`_wfn} is directly implemented
rather than \texttt{integrate\char`_sd\char`_sd}, then
\texttt{integrate\char`_sd\char`_wfn} can be vectorized over the given Slater
determinant and its excitations associated with the application of the
Hamiltonian.
When the derivative of the integral is not provided, orbital optimization is
only available through relatively inefficient gradient-free optimization algorithms, such as CMA-ES.

Since the H\"{u}ckel Hamiltonian is defined by its one-electron integrals, this
class can be used to describe any Hamiltonian with only one-body operators.
The integrals for the H\"{u}ckel Hamiltonian (and other model Hamiltonians) can
be generated using the ModelHamiltonian GitHub repository\cite{modelhamiltonian}.

\subparagraph{Implementing an Objective:}
New objectives can be implemented in \texttt{Fanpy} by making a subclass of the
objective base class.
The subclass requires the method \texttt{objective} to be defined.
To use the gradient (or Jacobian) in the optimization algorithm, the subclass
must also contain the method \texttt{gradient} (or \texttt{jacobian}).
These objectives can then be solved using the appropriate methods in the \texttt{solver} module. For example, the energy related objectives can be solved via minimization and the projected Schr\"{o}dinger equation related objectives can be solved via root-finding and least-squares algorithms.

For example, consider the local energy used in the
orbital-space variational Quantum Monte Carlo\cite{vqmc},

\begin{equation}
  \label{eq:fanpy_local_energy}
  E_L = \sum_{i} \frac{\braket{\Phi_i | \hat{H} | \Psi}}{\braket{\Phi_i | \Psi}}
\end{equation}
where the Slater determinant, $\Phi_i$, is sampled according to the distribution
$p(\Phi_i) = \frac{\braket{\Psi | \Phi_i}^2}{\sum_k \braket{\Psi | \Phi_j}^2}$. This corresponds to:
\begin{minted}{python}
from fanpy.eqn.base import BaseSchrodinger

class LocalEnergy(BaseSchrodinger):
    def __init__(self, wfn, ham, pspace, param_selection=None):
        super().__init__(wfn, ham, param_selection=param_selection)
        # param_selection is used to select the parameters that are active
        # throughout the optimization
        self.pspace = pspace
        # pspace is the list of Slater determinants from which local energy is
        # computed

    @property
    def num_eqns(self):
        # number of equations is used to differentiate objectives in the solver
        return 1

    def objective(self, params):
        # assign (active) parameters to the respective wavefunction and
        # Hamiltonian
        # note that params is always flattened (1-dimensional) for compatibility
        # with solvers
        self.assign_params(params)
        output = 0.0
        for sd in self.pspace:
            output += self.ham.integrate_sd_wfn(sd, self.wfn) / self.wfn.get_overlap(sd)
        return output

    def gradient(self, params):
        self.assign_params(params)
        # note that gradient of the objective is also flattened (1-dimensional)
        output = np.zeros(params.size)
        for sd in self.pspace:
            # indices of the wavefunction parameters that are active
            wfn_inds_component = self.indices_component_params[self.wfn]
            if wfn_inds_component.size > 0:
                # indices of the objective parameters that correspond to the
                # wavefunction
                wfn_inds_objective = self.indices_objective_params[self.wfn]

                # differentiate local energy with respect to wavefunction parameters
                output[wfn_inds_objective] += (
                    self.ham.integrate_sd_wfn(sd, self.wfn, wfn_deriv=wfn_inds_component)
                    / self.wfn.get_overlap(sd)
                )
                output[wfn_inds_objective] -= (
                    self.ham.integrate_sd_wfn(sd, self.wfn)
                    * self.wfn.get_overlap(sd, deriv=wfn_inds_component)
                    / self.wfn.get_overlap(sd) ** 2
                )
            # indices of the Hamiltonian parameters that are active
            # Used when hamiltonian has parameters to optimize (e.g. orbital
            # optimization)
            ham_inds_component = self.indices_component_params[self.ham]
            if ham_inds_component.size > 0:
                # indices of the objective parameters that correspond to the
                # hamiltonian
                ham_inds_objective = self.indices_objective_params[self.ham]

                # differentiate local energy with respect to Hamiltonian parameters
                output[ham_inds_objective] += (
                    self.ham.integrate_sd_wfn(sd, self.wfn, ham_deriv=ham_inds_component)
                    / self.wfn.get_overlap(sd)
                )
        return output
\end{minted}
Though it is not required, providing the indices in the gradient ensures that
users can specify the parameters that are active during the optimization via the
attribute \texttt{param\char`_selection}.

\section{Frequently Asked Questions}

\subparagraph{Who is \texttt{Fanpy} for?}
\texttt{Fanpy} was designed to be used by developers of post-HF methods,
especially those interested in new multireference wavefunction ans\"{a}tze.
Extensive programming experience is not necessary: \texttt{Fanpy}'s modular
design and extensive documentation make it easy to understand and extend the existing methods and base classes.
The base classes serve as templates to help ensure that the developed method
fits together with the rest of \texttt{Fanpy} seamlessly.
Developers with programming experience but a limited background in post-HF methods
should have an easier time understanding the code because the methods are
documented with the corresponding equations (and their derivations) and are
implemented in a simple and straight-forward fashion.

\subparagraph{What is the mission of \texttt{Fanpy}?}
Our goal is to develop a platform where developers of new \textit{ab initio} methods can
quickly implement and test their ideas. We hope to make it easier for researchers---whether they are seasoned professors or new graduate students---to test their ideas without being burdened by undocumented
code conventions, mysterious equations, or cumbersome installation processes.

\subparagraph{What does \texttt{Fanpy} do?}
As elaborated in Sections \ref{sec:fanpy_features} and \ref{sec:fanpy_examples}, \texttt{Fanpy}
provides independent modules that facilitate the development of new
multideterminant wavefunctions, Hamiltonians, representations of the Schr\"{o}dinger
equation (objective functions), and optimization algorithms.
We designed these modules to be compatible with one another so that researchers
can easily customize their calculations and experiment with different
combinations of methods and algorithms.

\subparagraph{What are the limits of \texttt{Fanpy}?}
At present, \texttt{Fanpy} is not designed for high performance.
In fact, its performance was often deliberately compromised to prioritize ease of use and
development. For accessibility, \texttt{Fanpy} was written in pure Python even though other languages, such as C++ and Julia, are often better suited for high-performance parallel computing.
Moreover, while \texttt{Fanpy}'s modular design is important for its extendibility and customizability, it prevents some types of algorithmic improvements. Since a method in its early stages of development is often intractably expensive, calculations in \texttt{Fanpy} are often limited to small model systems with small basis sets. Some of the more efficient methods (e.g. AP1roG, which could be extended to thousands of electrons in an efficient implementation) are limited to about 100 electrons in \texttt{Fanpy}. Consistent with the overall mission of HORTON 3, therefore, \texttt{Fanpy} should be viewed as a research tool that allows developers to quickly implement and test their ideas, rather than a
comprehensive quantum chemistry suite that can simulate large chemical systems. The intention is that after a researcher establishes that a method is of practical utility, a more efficient implementation can be developed.

\subparagraph{How do I install \texttt{Fanpy}?}
The \texttt{Fanpy} library can be installed directly from its source code available on
GitHub or through the \textit{pip} and \textit{conda} package-management systems.
Since \texttt{Fanpy} is purely Python and depends mainly on common Python
libraries (NumPy and SciPy), it can be installed by simply copying the source
code onto the desired directory (though this is not recommended).
For the most updated instructions on how to install \texttt{Fanpy}, please refer
to the \texttt{Fanpy} website.

\subparagraph{What is the future direction of \texttt{Fanpy}?}
In addition to developing additional methods relevant to our scientific interests, the next
iteration of \texttt{Fanpy} will focus on improving its performance.
Its computationally critical components will be outsourced to highly optimized
libraries, such as our in-house CI software, PyCI. Some of the performance bottlenecks will be removed by reimplementing some features in Cython or C++. As these improvements may cause problems for some users in terms of ease of use and installation, the pure Python implementation of \texttt{Fanpy} will
continue to be available.

In terms of features, modules for (arbitrary-order) perturbation theory, equations-of-motion, and quantum-mechanical embedding are in various stages of development.

\section{Summary}
This brief paper introduces \texttt{Fanpy} as a library for developing new
post-Hartree-Fock \textit{ab initio} methods.
\texttt{Fanpy}'s goal is to help researchers quickly test ideas for new correlated electronic structure theory methods and, to achieve this goal, \texttt{Fanpy} contains many methods that can be used in countless combinations
with one another. These methods are of intrinsic interest but, moreover, they serve as examples to be extended upon.
Base classes are available as templates to help users develop a structure that
is compatible with the rest of \texttt{Fanpy}.

\section*{Acknowledgements}

We wish to acknowledge various refinements to \texttt{Fanpy} library from
Matthew Chan, Cristina E. Gonz\'{a}lez-Espinoza, Xiaotian D. Yang, Stijn Fias,
Caitlin Lanssens, and the HORTON development team.

P.W.A. and F.H.Z. acknowledges Natural Sciences and Engineering Research Council (NSERC) of
Canada,Compute Canada, and CANARIE for financial
and computational support. P.W.A. acknowledge support from the Canada Research Chairs. T.V. acknowledges support from the Research Foundation – Flanders (FWO) (Belgium) and the Research Board of Ghent University. R.A.M.Q. acknowledges financial support from the University of Florida in the form
of a start-up grant.

\newpage
\printbibliography

@book{szabo,
author = {Szabo, A. and Ostlund, N.S.},
isbn = {0-486-69186-1},
pages = {43-107},
publisher = {McGraw-Hill Inc.},
title = {{Modern Quantum Chemistry - Introduction to Advanced Electronic Structure Theory}},
year = {1989},
}

@article{fci,
author = {Boys, S.F.},
journal = {Proceedings of the Royal Society of London A},
month = "2",
number = {1063},
pages = {542-554},
title = {{Electronic Wave Functions. I. A General Method of Calculation for the Stationary States of Any Molecular System}},
volume = {200},
year = {1950},
doi = {10.1098/rspa.1950.0036},
issn = {},
publisher = {},
url = {},
}

@article{sen0_ham,
author = {Richardson, R.W.},
journal = {Physical Review},
month = "7",
number = {4},
pages = {792-805},
title = {{Eigenstates of the $L=0$, Charge- and Spin-Independent Pairing Hamiltonian. I.
Seniority-Zero States}},
volume = {159},
year = {1967},
doi = {10.1103/PhysRev.159.792},
issn = {},
publisher = {},
url = {},
}

@article{hubbard_ham,
author = {Hubbard, J.},
journal = {Proceedings of the Royal Society of London A},
month = "11",
number = {1365},
pages = {238-257},
title = {{Electron Correlations in Narrow Energy Bands}},
volume = {276},
year = {1963},
doi = {10.1098/rspa.1963.0204},
issn = {},
publisher = {},
url = {},
}

@article{doci,
author = {Weinhold, F. and Wilson Jr, E.B.},
journal = {The Journal of Chemical Physics},
month = {},
number = {7},
pages = {2752-58},
title = {{Reduced Density Matrices of Atoms and Molecules. I. The 2 Matrix of Double-Occupancy,
Configuration-Interaction Wavefunctions for Singlet States}},
volume = {46},
year = {1967},
doi = {10.1063/1.1841109},
issn = {},
publisher = {},
url = {},
}

@article{cisd,
author = {Pople, J.A. and Seeger, R. and Krishnan, R.},
journal = {International Journal of Quantum Chemistry: Quantum Chemistry Symposium},
month = {},
number = {},
pages = {149-163},
title = {{Variational Configuration Interaction Methods and Comparison with Perturbation Theory}},
volume = {11},
year = {1977},
doi = {},
issn = {},
publisher = {},
url = {},
}

@article{cc_review,
author = {Bartlett, R.J. and Musia\l{}, M.},
journal = {Reviews of Modern Physics},
month = "2",
number = {1},
pages = {291-352},
title = {{Coupled-cluster theory in quantum chemistry}},
volume = {79},
year = {2007},
doi = {10.1103/RevModPhys.79.291},
issn = {},
publisher = {},
url = {},
}

@article{mps_dmrg,
author = {Schollw\"{o}ck, U.},
journal = {Annals of Physics},
month = "1",
number = {},
pages = {96-192},
title = {{The density-matrix renormalization group in the age of matrix product states}},
volume = {326},
year = {2011},
doi = {10.1016/j.aop.2010.09.012},
issn = {},
publisher = {},
url = {},
}

@article{geminal_first,
author = {Shull, H.},
journal = {The Journal of Chemical Physics},
title = {{Natural Spin Oribtal Analysis of Hydrogen Molecule Wave Functions}},
volume = {30},
number = {6},
pages = {1405-13},
year = {1959},
doi = {10.1063/1.1730212},
issn = {},
publisher = {},
url = {},
}

@article{apg,
author = {Kutzelnigg, W.},
journal = {The Journal of Chemical Physics},
title = {{Direct Determination of Natural Orbitals and Natural Expansion Coefficients of Many-Electron
Wavefunctions. I. Natural Orbitals in the Geminal Product Approximation}},
volume = {40},
number = {12},
pages = {3640-47},
year = {1964},
doi = {10.1063/1.1730212},
issn = {},
publisher = {},
url = {},
}

@article{apig_first,
author = {Silver, D.M.},
journal = {The Journal of Chemical Physics},
title = {{Natural Orbital Expansion of Interacting Geminals}},
volume = {50},
number = {12},
pages = {5108-16},
month = {},
year = {1969},
doi = {10.1063/1.1671025},
issn = {},
publisher = {},
url = {},
}

@article{apr2g,
author = {Johnson, P.A. and Ayers, P.W. and Limacher, P.A. and De Baerdemacker, S. and Van Neck, D. and Bultinck, P.},
journal = {Computational and Theoretical Chemistry},
title = {{A size-consistent approach to strongly correlated systems using a generalized
antisymmetrized product of nonorthogonal geminals}},
volume = {1003},
number = {},
pages = {101-13},
month = "1",
year = {2013},
doi = {10.1016/j.comptc.2012.09.030},
issn = {},
publisher = {},
url = {},
}

@article{ccsd,
author = {Purvis, G.D. and Bartlett R.J.},
journal = {The Journal of Chemical Physics},
month = "2",
number = {4},
pages = {1910-18},
title = {{A full coupled-cluster singles and doubles model: The inclusion of disconnected
triples}},
volume = {76},
year = {1982},
doi = {10.1063/1.443164},
issn = {},
publisher = {},
url = {},
}

@article{agp,
author = {Coleman, A.J.},
journal = {Journal of Mathematical Physics},
month = "9",
number = {9},
pages = {1425-1431},
title = {{Structure of Fermion Density Matrices. II. Antisymmeterized Geminal Powers}},
volume = {6},
year = {1965},
doi = {10.1063/1.1704794},
issn = {},
publisher = {},
url = {},
}

@book{1,
author = {Helgaker, T. and J{\o}rgensen, P. and Olsen, J.},
title = {{Modern electronic structure theory}},
year = {2000},
address = {Chichester},
publisher = {Wiley},
}

@article{2,
author = {Helgaker, T. and J{\o}rgensen, P. and Olsen, J.},
title = {{Modern electronic structure theory}},
journal = {Journal of Physical Chemistry},
year = {1996},
volume = {100},
pages = {13213-13225},
}

@article{7,
author = {Bytautas, L. and Henderson, T.M. and Jim\'{e}nez-Hoyos, C.A. and Ellis, J.K. and Scuseria, G.E.},
title = {{Seniority and orbital symmetry as tools for establishing a full configuration interaction hierarchy}},
journal = {Journal of Chemical Physics},
year = {2011},
volume = {135},
}

@article{10,
author = {Alcoba, D.R. and Torre, A. and Luis, L. and O\~{n}a, O.B. and Capuzzi, P. and Van Raemdonck, M. and Bultinck, P. and Van Neck, D.},
title = {{A hybrid configuration interaction treatment based on seniority number and excitation schemes}},
journal = {Journal of Chemical Physics},
year = {2014},
volume = {141},
issue = {24},
}

@article{15,
author = {Cook, D.B.},
title = {{Doubly-occupied orbital MCSCF methods}},
journal = {Molecular Physics},
year = {1975},
volume = {30},
pages = {733-743},
}

@article{16,
author = {Veillard, A. and Clementi, E.},
title = {{Complete multi-configuration self-consistent field theory}},
journal = {Theoretica Chimica Acta},
year = {1967},
volume = {7},
pages = {134-143},
}

@incollection{22,
author = {Paldus, J. and Li, X.Z.},
title = {{Critical assessment of coupled cluster method in quantum chemistry}},
booktitle = {Advances in Chemical Physics},
volume = {110},
editor = {Prigogine, I. and Rice, S.A.},
year = {1999},
pages = {1-175},
}

@article{23,
author = {Cizek, J.},
title = {{On the Correlation Problem in Atomic and Molecular Systems. Calculation of Wavefunction Components in Ursell-Type Expansion Using Quantum-Field Theoretical Methods}},
journal = {Journal of Chemical Physics},
year = {1966},
volume = {45},
pages = {4256-4266},
}

@book{24,
author = {Shavitt, I. and Bartlett, R.J.},
title = {{Many-body methods in chemistry and physics: MBPT and coupled-cluster theory}},
year = {2009},
address = {Cambridge},
publisher = {Cambridge},
}

@article{36,
author = {Wouters, S. and Poelmans, W. and De Baerdemacker, S. and Ayers, P.W. and Van Neck, D.},
title = {{CHEMPS2: Improved DMRG-SCF routine and correlation functions}},
journal = {Computer Physics Communications},
year = {2015},
volume = {191},
pages = {235-237},
}

@article{39,
author = {Hurley, A.C. and Lennard-Jones, J. and Pople, J.A.},
title = {{The molecular orbital theory of  chemical valency XVI}},
journal = {A theory of paired-electrons in polyatomic molecules Proceedings of the Royal Society of London Series A},
year = {1953},
volume = {220},
pages = {446-455},
}

@article{40,
author = {Parr, R.G. and Ellison, F.O. and Lykos, P.G.},
title = {{Generalized antisymmetric product wave functions for atoms and molecules}},
journal = {Journal of Chemical Physics},
year = {1956},
volume = {24},
pages = {1106},
}

@article{41,
author = {Parks, J.M. and Parr, R.G.},
title = {{Theory of Separated Electron Pairs}},
journal = {Journal of Chemical Physics},
year = {1958},
volume = {28},
pages = {335-345},
}

@article{42,
author = {McWeeny, R. and Sutcliffe, B.},
title = {{The density matrix in many-electron quantum mechancs III. Generalized product functions for Beryllium and Four-Electron Ions}},
journal = {Proceedings of the Royal Society of London Series A},
year = {1963},
volume = {273},
pages = {103-116},
}

@incollection{43,
author = {Surjan, P.R.},
title = {{An introduction to the theory of geminals}},
booktitle = {Correlation and Localization},
editor = {Surjan, P.R.},
year = {1999},
pages = {63-88},
}

@article{44,
author = {Allen, T.L. and Shull, H.},
title = {{Electron pairs in the Beryllium atom}},
journal = {Journal of Physical Chemistry},
year = {1962},
volume = {66},
pages = {2281-2283},
}

@article{45,
author = {Tecmer, P. and Boguslawski, K. and Johnson, P.A. and Limacher, P.A. and Chan, M. and Verstraelen, T. and Ayers, P.W.},
title = {{Assessing the Accuracy of New Geminal-Based Approaches}},
journal = {Journal of Physical Chemistry A},
year = {2014},
volume = {118},
issue = {39},
pages = {9058-9068},
}

@article{46,
author = {Paldus, J. and Cizek, J. and Sengupta, S.},
title = {{Geminal Localization in the Separated-Pair $\pi$-Electronic Model of Benzene}},
journal = {Journal of Chemical Physics},
year = {1971},
volume = {55},
issue = {5},
pages = {2452-2462},
}

@article{47,
author = {Paldus, J. and Sengupta, S. and Cizek, J.},
title = {{Diagrammatical Method for Geminals. II. Applications}},
journal = {Journal of Chemical Physics},
year = {1972},
volume = {57},
issue = {2},
pages = {652-666},
}

@article{48,
author = {Surjan, P.R. and Szabados, \'{A} and Jeszenszki, P. and Zoboki, T.},
title = {{Strongly orthogonal geminals: size-extensive and variational reference states}},
journal = {Journal of Mathematical Chemistry},
year = {2012},
volume = {50},
pages = {534-551},
}

@article{49,
author = {Rassolov, V.A.},
title = {{A geminal model chemistry}},
journal = {Journal of Chemical Physics},
year = {2002},
volume = {117},
pages = {5978-5987},
}

@article{50,
author = {Rassolov, V.A. and Xu, F. and Garashchuk, S.},
title = {{Geminal model chemistry II. Perturbative corrections}},
journal = {Journal of Chemical Physics},
year = {2004},
volume = {120},
pages = {10385-10394},
}

@article{51,
author = {Rassolov, V.A. and Xu, F.},
title = {{Geminal model chemistry III: Partial spin restriction}},
journal = {Journal of Chemical Physics},
year = {2007},
volume = {126},
pages = {234112},
}

@article{52,
author = {Cassam-Chena\"{i}, P.},
title = {{The electronic mean-field configuration interaction method. I. Theory and integral formulas}},
journal = {Journal of Chemical Physics},
year = {2006},
volume = {124},
pages = {194109},
}

@article{53,
author = {Cassam-Chena\"{i}, P. and Rassolov, V.},
title = {{The electronic mean field configuration interaction method: III - the $p$-orthogonality constraint}},
journal = {Chemical Physics Letters},
year = {2010},
volume = {487},
pages = {147-152},
}

@article{54,
author = {Cassam-Chena\"{i}, P. and Ilmane, A.},
title = {{Frequently asked questions on the mean field configuration interaction method. I-distinguishable degrees of freedom}},
journal = {Journal of Mathematical Chemistry},
year = {2012},
volume = {50},
pages = {652-667},
}

@article{55,
author = {Stein, T. and Henderson, T.M. and Scuseria, G.E.},
title = {{Seniority zero pair coupled cluster doubles theory}},
journal = {Journal of Chemical Physics},
year = {2014},
volume = {140},
issue = {21},
pages = {214113},
}

@article{56,
author = {Henderson, T.M. and Scuseria, G.E. and Dukelsky, J. and Signoracci, A. and Duguet, T.},
title = {{Quasiparticle coupled cluster theory for pairing interactions}},
journal = {Physical Review C},
year = {2014},
volume = {89},
issue = {5},
pages = {054305},
}

@article{57,
author = {Henderson, T.M. and Bulik, I.W. and Stein, T. and Scuseria, G.E.},
title = {{Seniority-based coupled cluster theory}},
journal = {Journal of Chemical Physics},
year = {2014},
volume = {141},
issue = {24},
pages = {244104},
}

@article{58,
author = {Bulik, I.W. and Henderson, T.M. and Scuseria, G.E.},
title = {{Can Single-Reference Coupled Cluster Theory Describe Static Correlation?}},
journal = {Journal of Chemical Theory and Computation},
year = {2015},
volume = {11},
issue = {7},
pages = {3171-3179},
}

@article{59,
author = {Cullen, J.},
title = {{Generalized valence bond solutions from a constrained coupled cluster method}},
journal = {Chemical Physics},
year = {1996},
volume = {202},
issue = {2-3},
pages = {217-229},
doi = {10.1016/0301-0104(95)00321-5},
}

@article{60,
author = {Miller, K.J. and Ruedenberg, K.},
title = {{Electron Correlation and Electron-Pair Wavefunction for the Beryllium Atom}},
journal = {Journal of Chemical Physics},
year = {1965},
volume = {43},
issue = {10},
pages = {S88-S90},
}

@article{61,
author = {Miller, K.J. and Ruedenberg, K.},
title = {{Electron Correlation and Separated-Pair Approximation. An Application to Berylliumlike Atomic Systems}},
journal = {Journal of Chemical Physics},
year = {1968},
volume = {48},
issue = {8},
pages = {3414-3443},
}

@article{62,
author = {Silver, D.M. and Mehler, E.L. and Ruedenberg, K.},
title = {{Electron Correlation and Separated Pair Approximation in Diatomic Molecules. I. Theory}},
journal = {Journal of Chemical Physics},
year = {1970},
volume = {52},
issue = {3},
pages = {1174-1180},
}

@article{63,
author = {Mehler, E.L. and Ruedenberg, K. and Silver, D.M.},
title = {{Electron Correlation and Separated Pair Approximation in Diatomic Molecules. II. Lithium Hydride and Boron Hydride}},
journal = {Journal of Chemical Physics},
year = {1970},
volume = {52},
issue = {3},
pages = {1181-1205},
}

@article{64,
author = {Silver, D.M. and Ruedenberg, K. and Mehler, E.L.},
title = {{Electron Correlation and Separated Pair Approximation in Diatomic Molecules. III. Imidogen}},
journal = {Journal of Chemical Physics},
year = {1970},
volume = {52},
issue = {3},
pages = {1206-1227},
}

@article{66,
author = {Coleman, A.J.},
title = {{The AGP model for fermion systems}},
journal = {International Journal of Quantum Chemistry},
year = {1997},
volume = {63},
issue = {1},
pages = {23-30},
}

@article{67,
author = {Bajdich, M. and Drobn\'{y}, G. and Wagner, L.K. and Schmidt, K.E.},
title = {{Pfaffian pairing wave functions in electronic-structure quantum Monte Carlo simulations}},
journal = {Physical Review Letters},
year = {2006},
volume = {96},
pages = {130201},
}

@article{68,
author = {Bajdich, M. and Mitas, L. and Wagner, L.K. and Schmidt, K.E.},
title = {{Pfaffian pairing and backflow wavefunctions for electronic structure quantum Monte Carlo methods}},
journal = {Physical Review B},
year = {2008},
volume = {77},
pages = {115112},
}

@article{69,
author = {Pernal, K.},
title = {{Intergeminal Correction to the Antisymmetrized Product of Strongly Orthogonal Geminals Derived from the Extended Random Phase Approximation}},
journal = {Journal of Chemical Theory and Computation},
year = {2014},
volume = {10},
issue = {10},
pages = {4332-4341},
}

@article{70,
author = {Pastorczak, E. and Pernal, K.},
title = {{ERPA-APSG: a computationally efficient geminal-based method for accurate description of chemical systems}},
journal = {Physical Chemistry Chemical Physics},
year = {2015},
volume = {17},
issue = {14},
pages = {8622-8626},
}

@article{71,
author = {Limacher, P.A. and Kim, T.D. and Ayers, P.W. and Johnson, P.A. and De Baerdemacker, S. and Van Neck, D.},
title = {{The influence of orbital rotation on the energy of closed-shell wavefunctions}},
journal = {Molecular Physics},
year = {2014},
volume = {112},
issue = {5-6},
pages = {853-862},
}

@article{72,
author = {Limacher, P.A.},
title = {{A new wavefunction hierarchy for interacting geminals}},
journal = {Journal of Chemical Physics},
year = {2016},
volume = {145},
issue = {19},
pages = {194102},
}

@article{73,
author = {Johnson, P.A. and Limacher, P.A. and Kim, T.D. and Richer, M. and Miranda-Quintana, R.A. and Heidar-Zadeh, F. and Ayers, P.W. and Bultinck, P. and De Baerdemacker, S. and Van Neck, D.},
title = {{Strategies for extending geminal-based wavefunctions: Open shells and beyond}},
journal = {Computational and Theoretical Chemistry},
year = {2017},
volume = {1116},
pages = {207-219},
}

@article{74,
author = {Boguslawski, K. and Tecmer, P. and Ayers, P.W. and Bultinck, P. and De Baerdemacker, S. and Van Neck, D.},
title = {{Efficient description of strongly correlated electrons with mean-field cost}},
journal = {Physical Review B},
year = {2014},
volume = {89},
issue = {20},
pages = {201106},
}

@article{75,
author = {Boguslawski, K. and Tecmer, P. and Bultinck, P. and De Baerdemacker, S. and Van Neck, D. and Ayers, P.W.},
title = {{Nonvariational Orbital Optimization Techniques for the AP1roG Wave Function}},
journal = {Journal of Chemical Theory and Computation},
year = {2014},
volume = {10},
issue = {11},
pages = {4873-4882},
}

@article{ap1rog,
author = {Limacher, P.A. and Ayers, P.W. and Johnson, P.A. and De Baerdemacker, S. and Van Neck, D. and Bultinck, P.},
journal = {Journal of Chemical Theory and Computation},
title = {{A New Mean-Field Method Suitable for Strongly Correlated Electrons: Computationally Facile Antisymmetric Products of Nonorthogonal Geminals}},
volume = {9},
number = {3},
pages = {1394-1401},
month = {},
year = {2013},
doi = {10.1021/ct300902c},
issn = {},
publisher = {},
url = {},
}

@article{ramonemail_cc1,
author = {Evangelista, F. A. and Chan, G. K. L. and Scuseria, G. E.},
title = {Exact parameterization of fermionic wave functions via unitary coupled cluster theory},
journal = {Journal of Chemical Physics},
year = {2019},
volume = {151},
pages = {244112},
}

@article{ramonemail_cc2,
author = {Evangelista, F. A.},
title = {Alternative single-reference coupled cluster approaches for multireference problems: The simpler, the better},
journal = {Journal of Chemical Physics},
year = {2011},
volume = {134},
pages = {224102},
}

@misc{gaussian,
author={M. J. Frisch and G. W. Trucks and H. B. Schlegel and G. E. Scuseria and M. A. Robb and J. R. Cheeseman and G. Scalmani and V. Barone and G. A. Petersson and H. Nakatsuji and X. Li and M. Caricato and A. V. Marenich and J. Bloino and B. G. Janesko and R. Gomperts and B. Mennucci and H. P. Hratchian and J. V. Ortiz and A. F. Izmaylov and J. L. Sonnenberg and D. Williams-Young and F. Ding and F. Lipparini and F. Egidi and J. Goings and B. Peng and A. Petrone and T. Henderson and D. Ranasinghe and V. G. Zakrzewski and J. Gao and N. Rega and G. Zheng and W. Liang and M. Hada and M. Ehara and K. Toyota and R. Fukuda and J. Hasegawa and M. Ishida and T. Nakajima and Y. Honda and O. Kitao and H. Nakai and T. Vreven and K. Throssell and Montgomery, {Jr.}, J. A. and J. E. Peralta and F. Ogliaro and M. J. Bearpark and J. J. Heyd and E. N. Brothers and K. N. Kudin and V. N. Staroverov and T. A. Keith and R. Kobayashi and J. Normand and K. Raghavachari and A. P. Rendell and J. C. Burant and S. S. Iyengar and J. Tomasi and M. Cossi and J. M. Millam and M. Klene and C. Adamo and R. Cammi and J. W. Ochterski and R. L. Martin and K. Morokuma and O. Farkas and J. B. Foresman and D. J. Fox},
title={Gaussian 16 {R}evision {C}.01},
year={2016},
note={Gaussian Inc. Wallingford CT}
}

@inbook{cma,
author="Hansen, N. and Auger, A.",
editor="Borenstein, Y. and Moraglio, A.",
title="Principled Design of Continuous Stochastic Search: From Theory to Practice",
bookTitle="Theory and Principled Methods for the Design of Metaheuristics",
year="2014",
publisher="Springer Berlin Heidelberg",
address="Berlin, Heidelberg",
pages="145-180",
isbn="978-3-642-33206-7",
doi="10.1007/978-3-642-33206-7_8",
url="https://doi.org/10.1007/978-3-642-33206-7_8"
}

@article{vqmc4,
  title={The Jastrow antisymmetric geminal power in Hilbert space: Theory, benchmarking, and application to a novel transition state},
  author={Neuscamman, Eric},
  journal={The Journal of chemical physics},
  volume={139},
  number={19},
  pages={194105},
  year={2013},
  publisher={American Institute of Physics}
}

@article{vqmc7,
  title={Improved optimization for the cluster Jastrow antisymmetric geminal power and tests on triple-bond dissociations},
  author={Neuscamman, Eric},
  journal={Journal of chemical theory and computation},
  volume={12},
  number={7},
  pages={3149--3159},
  year={2016},
  publisher={ACS Publications}
}

@article{vqmc8,
  title={Variational Monte Carlo method in the presence of spin-orbit interaction and its application to Kitaev and Kitaev-Heisenberg models},
  author={Kurita, Moyuru and Yamaji, Youhei and Morita, Satoshi and Imada, Masatoshi},
  journal={Physical Review B},
  volume={92},
  number={3},
  pages={035122},
  year={2015},
  publisher={APS}
}

@book{qmc2,
  title={Quantum Monte Carlo methods in physics and chemistry},
  author={Nightingale, M Peter and Umrigar, Cyrus J},
  number={525},
  year={1998},
  publisher={Springer Science \& Business Media}
}

@article{molpro,
  title={Molpro: a general-purpose quantum chemistry program package},
  author={Werner, Hans-Joachim and Knowles, Peter J and Knizia, Gerald and Manby, Frederick R and Sch{\"u}tz, Martin},
  journal={Wiley Interdisciplinary Reviews: Computational Molecular Science},
  volume={2},
  number={2},
  pages={242--253},
  year={2012},
  publisher={Wiley Online Library}
}

@misc{horton,
    author       = {Verstraelen, Toon and Tecmer, Pawel and Heidar-Zadeh, Farnaz and Boguslawski, Katharina and Chan, Matthew and Zhao, Yilin and Kim, Taewon D. and Vandenbrande, Steven and Yang, Derrick and Gonz\'{a}lez-Espinoza, Cristina E. and Fias, Stijn and Limacher, Peter A. and Berrocal, Diego and Malek, Ali and Ayers, Paul W.},
    title        = {{HORTON}},
    year         = 2015,
    version      = {2.0.1},
    url          = {http://theochem.github.com/horton/}
    }

@article{pyscf,
  title={PySCF: the Python-based simulations of chemistry framework},
  author={Sun, Qiming and Berkelbach, Timothy C and Blunt, Nick S and Booth, George H and Guo, Sheng and Li, Zhendong and Liu, Junzi and McClain, James D and Sayfutyarova, Elvira R and Sharma, Sandeep and others},
  journal={Wiley Interdisciplinary Reviews: Computational Molecular Science},
  volume={8},
  number={1},
  pages={e1340},
  year={2018},
  publisher={Wiley Online Library}
}

@article{psi4,
  title={PSI4 1.4: Open-source software for high-throughput quantum chemistry},
  author={Smith, Daniel GA and Burns, Lori A and Simmonett, Andrew C and Parrish, Robert M and Schieber, Matthew C and Galvelis, Raimondas and Kraus, Peter and Kruse, Holger and Di Remigio, Roberto and Alenaizan, Asem and others},
  journal={The Journal of Chemical Physics},
  volume={152},
  number={18},
  pages={184108},
  year={2020},
  publisher={AIP Publishing LLC}
}

@article{psi4numpy,
  title={Psi4NumPy: An interactive quantum chemistry programming environment for reference implementations and rapid development},
  author={Smith, Daniel GA and Burns, Lori A and Sirianni, Dominic A and Nascimento, Daniel R and Kumar, Ashutosh and James, Andrew M and Schriber, Jeffrey B and Zhang, Tianyuan and Zhang, Boyi and Abbott, Adam S and others},
  journal={Journal of chemical theory and computation},
  volume={14},
  number={7},
  pages={3504--3511},
  year={2018},
  publisher={ACS Publications}
}

@article{vqmc,
  title={Improved speed and scaling in orbital space variational Monte Carlo},
  author={Sabzevari, Iliya and Sharma, Sandeep},
  journal={Journal of chemical theory and computation},
  volume={14},
  number={12},
  pages={6276--6286},
  year={2018},
  publisher={ACS Publications}
}

@article{scipy,
  title={SciPy 1.0: fundamental algorithms for scientific computing in Python},
  author={Virtanen, Pauli and Gommers, Ralf and Oliphant, Travis E and Haberland, Matt and Reddy, Tyler and Cournapeau, David and Burovski, Evgeni and Peterson, Pearu and Weckesser, Warren and Bright, Jonathan and others},
  journal={Nature methods},
  volume={17},
  number={3},
  pages={261--272},
  year={2020},
  publisher={Nature Publishing Group}
}

@misc{pycma,
  author       = {Nikolaus Hansen and Youhei Akimoto and Petr Baudis},
  title        = {{CMA-ES/pycma} on {G}ithub},
  howpublished = {Zenodo, DOI:10.5281/zenodo.2559634},
  month        = feb,
  year         = 2019,
  doi          = {10.5281/zenodo.2559634},
  url          = {https://doi.org/10.5281/zenodo.2559634},
}

@software{scikit,
  author       = {Tim Head and
                  MechCoder and
                  Gilles Louppe and
                  Iaroslav Shcherbatyi and
                  fcharras and
                  Z\'{e} Vin\'{i}cius and
                  cmmalone and
                  Christopher Schr\"{o}der and
                  nel215 and
                  Nuno Campos and
                  Todd Young and
                  Stefano Cereda and
                  Thomas Fan and
                  rene-rex and
                  Kejia (KJ) Shi and
                  Justus Schwabedal and
                  carlosdanielcsantos and
                  Hvass-Labs and
                  Mikhail Pak and
                  SoManyUsernamesTaken and
                  Fred Callaway and
                  Lo\"{i}c Est\`{e}ve and
                  Lilian Besson and
                  Mehdi Cherti and
                  Karlson Pfannschmidt and
                  Fabian Linzberger and
                  Christophe Cauet and
                  Anna Gut and
                  Andreas Mueller and
                  Alexander Fabisch},
  title        = {scikit-optimize/scikit-optimize: v0.5.2},
  month        = mar,
  year         = 2018,
  publisher    = {Zenodo},
  version      = {v0.5.2},
  doi          = {10.5281/zenodo.1207017},
  url          = {https://doi.org/10.5281/zenodo.1207017}
}

@article{ppp1,
  title={A Semi-Empirical Theory of the Electronic Spectra and Electronic Structure of Complex Unsaturated Molecules. I.},
  author={Pariser, Rudolph and Parr, Robert G},
  journal={The Journal of Chemical Physics},
  volume={21},
  number={3},
  pages={466--471},
  year={1953},
  publisher={American Institute of Physics}
}

@article{ppp2,
  title={A semi-empirical theory of the electronic spectra and electronic structure of complex unsaturated molecules. II},
  author={Pariser, Rudolph and Parr, Robert G},
  journal={The Journal of Chemical Physics},
  volume={21},
  number={5},
  pages={767--776},
  year={1953},
  publisher={American Institute of Physics}
}

@article{ppp3,
  title={Electron interaction in unsaturated hydrocarbons},
  author={Pople, John A},
  journal={Transactions of the Faraday Society},
  volume={49},
  pages={1375--1385},
  year={1953},
  publisher={Royal Society of Chemistry}
}

@book{huckel,
  title={Second quantized approach to quantum chemistry: an elementary introduction},
  author={Surj{\'a}n, P{\'e}ter R},
  year={2012},
  publisher={Springer Science \& Business Media}
}

@article{ising,
  title={Beitrag zur theorie des ferromagnetismus},
  author={Ising, Ernst},
  journal={Zeitschrift f{\"u}r Physik},
  volume={31},
  number={1},
  pages={253--258},
  year={1925},
  publisher={Springer}
}

@incollection{heisenberg,
  title={Zur theorie des ferromagnetismus},
  author={Heisenberg, Werner},
  booktitle={Original Scientific Papers Wissenschaftliche Originalarbeiten},
  pages={580--597},
  year={1985},
  publisher={Springer}
}

@article{models,
  title={Density functionals and model Hamiltonians: Pillars of many-particle physics},
  author={Capelle, Klaus and Campo Jr, Vivaldo L},
  journal={Physics Reports},
  volume={528},
  number={3},
  pages={91--159},
  year={2013},
  publisher={Elsevier}
}

@article{richardson1,
  title={A restricted class of exact eigenstates of the pairing-force Hamiltonian},
  author={Richardson, RW},
  year={1963},
  publisher={Elsevier}
}

@article{richardson2,
  title={Colloquium: Exactly solvable Richardson-Gaudin models for many-body quantum systems},
  author={Dukelsky, J and Pittel, S and Sierra, G},
  journal={Reviews of modern physics},
  volume={76},
  number={3},
  pages={643},
  year={2004},
  publisher={APS}
}

@article{huckel2,
  title={Semiempirical Hamiltonians for spatially confined $\pi$-electron systems},
  author={Planelles, J and Zicovich-Wilson, C and Jaskolski, W and Corma, A},
  journal={International journal of quantum chemistry},
  volume={60},
  number={5},
  pages={971--981},
  year={1996},
  publisher={Wiley Online Library}
}

@book{jensen,
  title={Introduction to computational chemistry},
  author={Jensen, Frank},
  year={2017},
  publisher={John wiley \& sons}
}

@book{cramer,
  title={Essentials of computational chemistry: theories and models},
  author={Cramer, Christopher J},
  year={2013},
  publisher={John Wiley \& Sons}
}

@book{piela,
  title={Ideas of quantum chemistry},
  author={Piela, Lucjan},
  year={2013},
  publisher={Elsevier}
}

@article{qmc,
  title={Observations on variational and projector Monte Carlo methods},
  author={Umrigar, CJ},
  journal={The Journal of chemical physics},
  volume={143},
  number={16},
  pages={164105},
  year={2015},
  publisher={AIP Publishing LLC}
}

@unpublished{fanci,
author = {Kim, T. D. and Miranda-Quintana, R. A. and Richer, M. and Ayers, P. W.},
title = {Flexible Ansatz for N-body Configuration Interaction},
note = {Unpublished Manuscript},
year = 2020,
}

@unpublished{fanpy,
author = {Kim, T. D. and Richer, M. and S\'{a}nchez-D\'{i}az, G. and Heidar-Zadeh, F. and Verstraelen, T. and Miranda-Quintana, R. A. and Ayers, P. W.},
title = {Fanpy},
note = {Unpublished Manuscript},
year = 2020,
}

@misc{modelhamiltonian,
    author       = {Adams, Wil and Zhao, Yilin and Ayers, Paul W.},
    title        = {{ModelHamiltonian}},
    year         = 2020,
    version      = {0.0.0},
    url          = {https://github.com/QuantumElephant/ModelHamiltonian}
    }

% Submissions are not required to reflect the precise reference formatting of
% the journal (use of italics, bold etc.), however it is important that all key
% elements of each reference are included.
% \bibliography{references}

\end{document}